\documentclass[10pt]{iopart}
\usepackage{array,graphicx,float,epsfig,wrapfig}
\usepackage{amssymb}
\usepackage{multirow,booktabs}
\usepackage[nooneline,scriptsize,bf]{caption2}

\begin{document}

\title[Improving ellipticity detection ...]{Improving ellipticity detection sensitivity for the Q \& A vacuum birefringence experiment}

\author[S-J Chen, H-H Mei, W-T Ni and J-S Wu ]{Sheng-Jui Chen, Hsien-Hao Mei, Wei-Tou Ni and Jeah-Sheng Wu}

\address{Center for Gravitation and Cosmology, Department of Physics,\\
National Tsing Hua University,
%101 Kuang Fu Rd., Sec.2
Hsinchu, Taiwan 30055, Republic of China}

\begin{abstract}
Q \& A (quantum electrodynamics test and search for axion) experiment was first proposed in 1994 and
a 3.5 m high-finesse Fabry-Perot prototype detector extendable to 7 m has been built and tested.
We use X-pendulums and automatic control schemes developed by the gravitational-wave detection community
for mirror suspension and cavity control.
In this paper, we first give an overview of ellipticity detection scheme of Q \& A experiment
and compare it with other experiments in which Fabry-Perot cavities are also used to multiply the effectiveness
of the magnetic field.  We then present the displacement spectra of our suspension system and use the data for designing
automatic alignment control.
Results and experiences gained in the previous test runs have pointed out that the lateral
and rotational motion of our suspension system could cause noises and laser off-locks from the Fabry-Perot cavity,
and downgrade the ellipticicy detection sensitivity.
To fix this problem, i.e., to maintain the alignment between cavity mirrors and incoming laser beam, and to minimize detection
noises, we implement an automatic alignment control scheme based on the Ward differential wavefront sensing technique.
To improve ellipsometry resolution, we use new polarizers of extinction ratio smaller than $10^{-8}$.
\end{abstract}

\ead{d883374@oz.nthu.edu.tw}
\pacs{04.80.-y, 12.20.-m, 14.80.Mz, 07.60.Ly, 07.60.Fs, 33.55.Ad}


%\maketitle

\section{Introduction}
Quantum Electrodynamics (QED) predicts that vacuum is birefringent in a transverse magnetic field.  For a B field of 2.5 T, the birefringence index $\Delta n\!\equiv\!n_{\parallel}-n_{\perp}$ ($\propto\!B^2$) is $2.5\!\times\!10^{-23}$.
Iacopini and Zavattini [1] proposed to measure
the vacuum birefringence in a 10 T magnetic field in the laboratory by a precision
determination of the induced ellipticity on a laser beam down to 10$^{-11}$.
The development of ultrahigh-precision technology  in the
laser-interferometric gravitational-wave detection community prompted our thought of its application to this matter [2].
Cameron {\it et al.} [3] performed the first precision measurement of induced
ellipticity and polarization rotation of light in vacuum in a magnetic field using a multipass cavity.
The result is good to put a
limit on the axion-photon coupling; however, it is about 3
orders short from measuring QED birefringence effect.
\begin{figure}
\begin{center}
\includegraphics[scale=0.7]{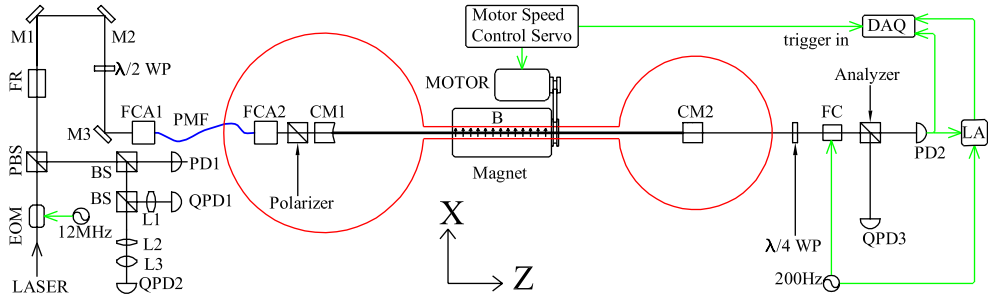}
\caption{\scriptsize Experimental Setup. EOM electro-optical modulator;
$\lambda/2$ WP half-wave plate; $\lambda/4$ WP quarter-wave plate; L1, L2, L3 lenses; M1, M2, M3 reflection mirrors;
 PBS polarizing beam splitter; BS beam spiltter; FR Faraday rotator; FCA1, FCA2 fibre coupler assemblies; PMF polarization maintaining
fibre;  CM1, CM2 cavity mirrors; B magnetic field; PD1, PD2 phtodetectors; QPD1, QPD2, QPD3 quadrant photodiodes
; FC Faraday cell; DAQ data acquisition system; LA lock-in amplifier.}
\end{center}
\end{figure}
In 1994, 3  experiments were put forward
 and  started for measuring the vacuum berefringence:
the PVLAS experiment [4], the Fermilab P-877 experiment  [5], and  the Q \& A experiment [6].
In 1998, we presented a comprehensive comparison of figures of merit of the 3 experiments [6].
PVLAS is now well underway [7].
Fermilab P-877 experiment was terminated in 2000.
We have constructed and tested our prototype 3.5 m
high-finesse Fabry-Perot inteferometer with ellipsometry [8].  After this first phase, we entered the second
phase of improving the detection sensitivity. Figure 1 shows the experimental setup.
The laser beam is modulated by a 12 MHz electro-optical modulator (EOM) and injected via a fibre coupler assembly (FCA1),
a polarizing maintaining fibre (PMF), the FCA2 on the mirror suspension system, and a Glan-Taylor
polarizer to the cavity mirror CM1.
The PMF serves as
\begin{wrapfigure}[15]{l}{0.5\textwidth}
%\begin{center}
%\centering
\includegraphics[scale=0.44]{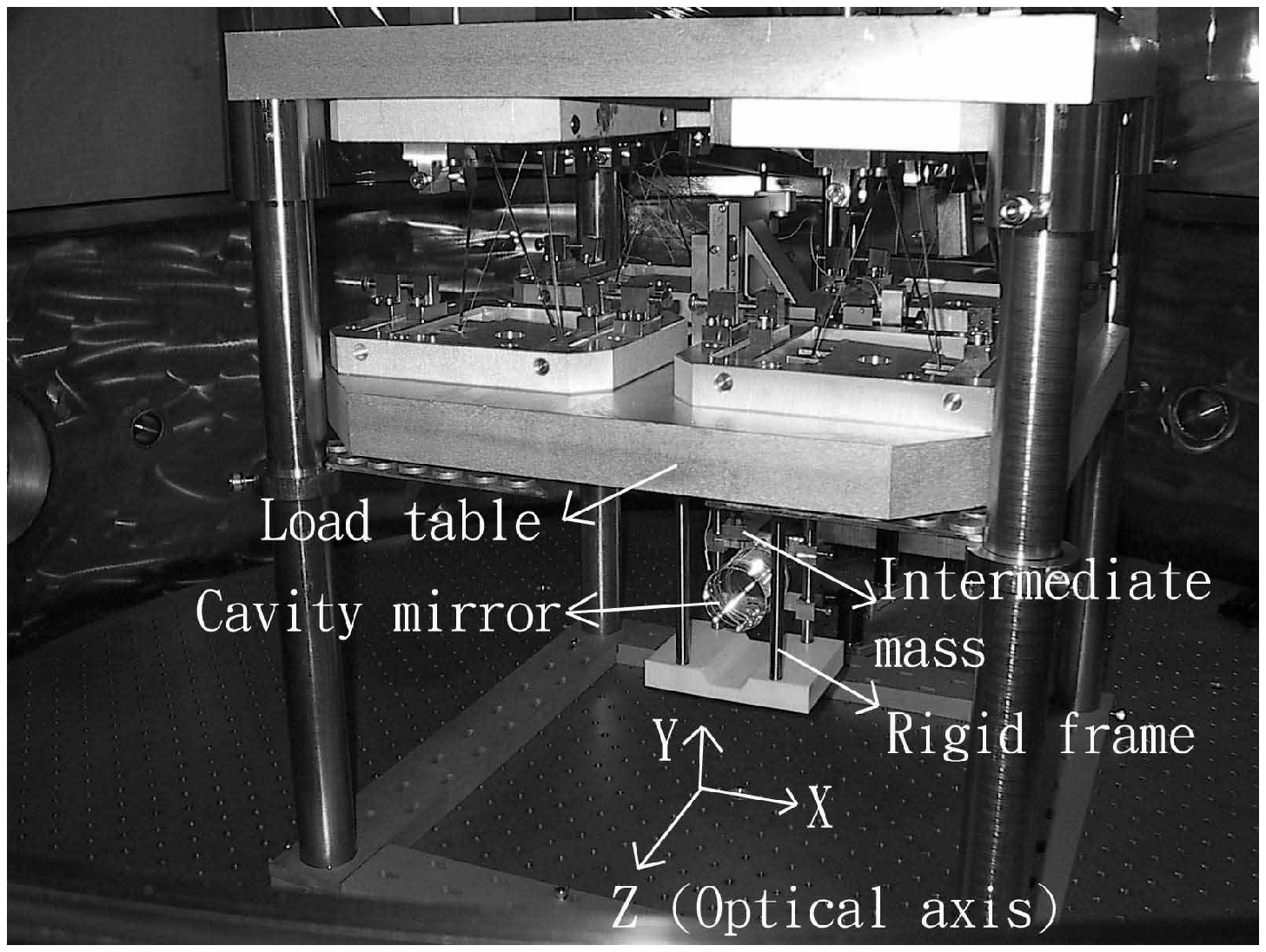}
%\includegraphics[angle=90,width=0.45\textwidth,height=0.3\textwidth]{x_pic2.eps}
%\caption{\scriptsize Picture of the suspension system.}
%\end{center}
\mbox{\scriptsize ~~~~~~~~{\bf Figure 2.} Picture of the suspension system.}
\end{wrapfigure}
 a simple mode cleaner. The laser is stabilized to the 3.5 m cavity by the Pound-Drever-Hall method.
When we achieve the second-phase goals, we will start
the third phase of extending the interferometer to 7 m and the magnet to 5 m to measure the vacuum birefringence.
A comparison of the PVLAS experiment [7], the new BMV
 experiment [9] and the Q \& A experiment
in their developments is compiled in Table 1.  With the goals achieved, the expected measurement times to QED vacuum birefringence
level for PVLAS experiment, BMV experiment and Q \& A experiment are one day, 10 pulses and 2 days respectively.
\begin{table}
\caption{{\scriptsize A Compilation of the Ongoing QED Vacuum Birefringence Experiments --- PVLAS, BMV and Q\&A Experiment.}}
\begin{center}
\fontsize{7}{10pt}\selectfont
\begin{tabular}{@{}lccc@{}}
\toprule
Experiment & PVLAS & BMV & Q \& A Experiment\\
\midrule
{\bf Status} & Achieved/{\it Goal} & Achieved/{\sf 2-year goal}/{\it Goal} & Achieved/{\sf 2-year goal}/{\it Goal}\\

Wavelength $\lambda$ (nm) & 1064 & 1064 & 1064/{\sf 1064}/{\it 532}\\

Type of dipole & Rotating & \multirow{2}{*}[1pt]{Pulsed}
& \multirow{2}{*}[1pt]{Switching/}{\sf Rotating}\\[-2pt] magnet & superconducting &  & ~~~~~~~~~~~~~~~~{\sf permanent}\\

Square of magnetic & \multirow{2}{*}[1pt]{40} & \multirow{2}{*}[1pt]{36/{\sf 100}/{\it 625}} & \multirow{2}{*}[1pt]{0.72/{\sf 5.3}/{\it 6.25}}\\[-2pt] field $B^2$
(T$^2$)\\

Length of magnetic & \multirow{2}{*}[1pt]{1} & \multirow{2}{*}[1pt]{0.3/{\sf 0.3}/{\it 1.5}} & \multirow{2}{*}[1pt]{0.2/{\sf 0.6}/{\it
5}}\\[-2pt] field $L_B$ (m)\\

Finesse of Fabry- & \multirow{2}{*}[1pt]{120,000} & \multirow{2}{*}[1pt]{50,000/{\sf 200,000}/{\it 1,000,000}} & \multirow{2}{*}[1pt]{12,000/{\sf 20,000}/{\it
100,000}}\\[-2pt] Perot cavity $F$\\

Modulation frequency & \multirow{3}{*}[1pt]{1} & \multirow{3}{*}[1pt]{37}
& \multirow{3}{*}[1pt]{0.05/{\sf 10}/{\it 10}}\\[-2pt] of magnetic field $f_m$ & & & \\[-2pt] (Hz) or 1/duration\\

%Easiness of vibration &
%\multirow{3}{*}[1pt]{400} & \multirow{3}{*}[1pt]{$\sim\!550,000$} & \multirow{3}{*}[1pt]{1}
%& \multirow{3}{*}[1pt]{40,000}\\[-2pt] isolation at the field &&&&\\[-2pt] modulation frequency ($\propto\! f_m^{-2}$)\\

Duration of continuous & \multirow{2}{*}[1pt]{4 hr} & \multirow{2}{*}[1pt]{100 msec} &
\multirow{2}{*}[1pt]{$>$10 days}\\[-2pt] magnet operation $T_1$\\

Magnetic regeneration & \multirow{2}{*}[1pt]{$4\!\sim\!6 $ hr} & \multirow{2}{*}[1pt]{15 min} &
\multirow{2}{*}[1pt]{$<$10 min}\\[-2pt] time $T_2$\\

Magnetic duty cycle $R$ & $0.4\!\sim\!0.5$ & $\sim 10^{-4}$ & $\sim 1$\\

Effective magnetic factor & \multirow{2}{*}[1pt]{$25.3\!\sim\!28.3$} & \multirow{2}{*}[1pt]{0.1/{\sf 0.3}/{\it 10}} &
\multirow{2}{*}[1pt]{0.17/{\sf 3}/{\it 30}}\\[-2pt] $\sim\!B^2\!L_B\!R^{1/2}$ (T$^2\,\cdot$ m)\\

Sensitivity (rad$\cdot$Hz$^{-1/2}$) & $10^{-7}$/${\it 10^{-8}}$ & ${\it 1\!\times\!10^{-9}}$ & $5\!\times\!10^{-6}$/${\sf 5\!\times\!10^{-8}}$/${\it 1\!\times\!10^{-8}}$\\
Generated vacuum & \multirow{2}{*}[1pt]{$3.2\!\times\!10^{-11}$} & \multirow{2}{*}[1pt]{${\it 7\!\times\!10^{-9}}$}
& \multirow{2}{*}[1pt]{$10^{-14}$/${\sf 5\!\times\!10^{-13}}$/${\it 2.3\!\times\!10^{-11}}$}\\[-2pt] birefringent effect (rad)\\
\bottomrule
\end{tabular}
\end{center}
\label{table1}
\end{table}
In the following, we present the test results of our X-pendulum-double-pendulum suspension system, scheme of automatic alignment control  and method
of improving ellipsometry.  In [10], we will present the magnetic profile of our prototype rotatable permanent magnet and discuss heterodyne detection schemes.

\section{Suspension system}

Our suspension system for a 3.5 m-cavity mirror consists of an X-pendulum-double-pendulum set mounted on an isolated table
inside a vacuum chamber fixed to ground using bellows (Figure 2).
X-pendulum, named after the crossed-wire structure used in it, was originally designed by Barton, Kuroda, Tatsumi and Uchiyama [11] of
the Institute for Cosmic Ray Research, University of Tokyo, and
planned to be installed in TAMA 300 project.  We follow closely their design.
The X-pendulum has its resonant frequency around 0.28 Hz (Figure 3) and attenuates
the horizontal seismic noise in frequency range between 1-20 Hz.
The Fabry-Perot cavity mirror is attached to the load table of the X-pendulum via a
double pendulum mechanism.
%, where the intermediate mass has been modified to be
%able to accommodate two magnets for control of the cavity mirror.
To test our suspension system, we
have measured the background seismic noise and the relative displacement control noise between
two cavity mirrors using heterodyne interferometry.  For seismic noise measurement, we simply measured the displacement
of the suspended cavity mirror relative to a reference point on the optical table on which the
suspension system is mounted; the measured result is the resonant displacement of the suspension
itself relative to the seismic motion on the optical bench. For relative displacement between cavity mirrors,
a diode-pumped Nd:YAG laser was frequency stabilized to the suspended
Fabry-Perot cavity via the Pound-Drever-Hall locking technique.  The relative displacement ($>$ 1 Hz) is
estimated from the correction signal that drives the PZT of the laser.  The
spectral density of these displacements are plotted in figure 3, it shows that our suspension system
is capable of isolating seismic noise in frequency range between 1-20 Hz.
There are many small resonant modes in this frequency band; however, we can
avoid these peaks by choosing suitable modulation frequency of the magnetic field.  The noise floor at 5-10 Hz
is 2-order of magnitude below that at 0.05 Hz.
%\begin{minipage}[t]{0.5\textwidth}
%\begin{figure}
%\begin{center}
%\centering
%\includegraphics{x_pic.eps}
%\includegraphics[angle=90,scale=0.4]{x_pic2.eps}
%\caption{\scriptsize Picture of the suspension system.}
%\end{center}
%\end{figure}
%\end{minipage}
%\begin{figure}
%\begin{center}
%\includegraphics[scale=0.5]{disp_noises2.eps}
%\caption{\scriptsize Displacement noises. Curve 1 and 2:  Combined displacement of the seismic
%noises and suspension's resonant motions in Z and X axes respectively. Curve 3:  Relative
%displacement between CM1 and CM2 in Z axis.  Curve 4:  Simulation result of the automatic alignment
% control where curve 2 was used as the noise source.}
%\end{center}
%\end{figure}

\section{Scheme of automatic alignment system in Q \& A experiment}

  From the test result of our suspension system, we know that the X-pendulum has a large rms
amplitude (3-5 $\mu$m) in 2 horizontal directions of displacement.  We need either to tune the frequency of the laser
to the cavity length change or to apply force to the cavity mirrors to adjust their
positions by a control servo.  In the direction parallel to the cavity axis, this has been successfully done
in previous test runs [8].  The
lateral and rotational motions of the suspension systems misalign the cavity with respect to
\begin{wrapfigure}[19]{r}{0.5\textwidth}
%\begin{center}
\includegraphics[scale=0.36]{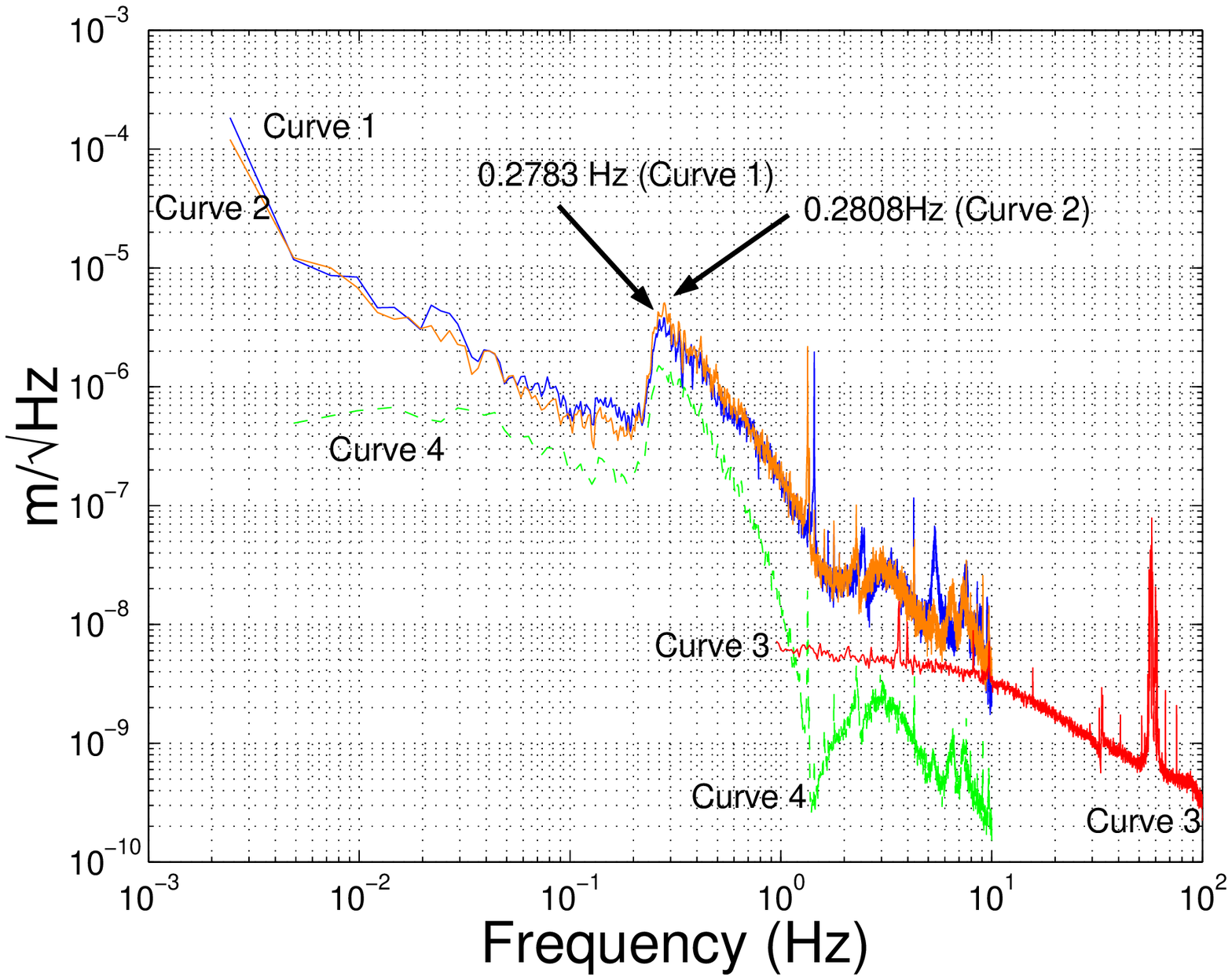}
%\includegraphics[width=0.45\textwidth,height=0.4\textwidth]{disp_noises2.eps}
%\caption{\scriptsize Displacement noises. Curve 1 and 2:  Combined displacement of the seismic
%noises and suspension's resonant motions in Z and X axes respectively. Curve 3:  Relative
%displacement between CM1 and CM2 in Z axis.  Curve 4:  Simulation result of the automatic alignment
% control where curve 2 was used as the noise source.}
%\end{center}
%\mbox{\scriptsize {\bf Figure 3.} Displacement noises. Curve 1 and 2:}
%\mbox{\scriptsize Combined~ displacement~ of~ the~ seismic~ noises}
%\mbox{\scriptsize and suspension's resonant motions in Z and X axes}
%\mbox{\scriptsize respectively. Curve 3:  Relative displacement}
%\mbox{\scriptsize between CM1 and CM2 in Z axis. Curve 4:}
%\mbox{\scriptsize Simulation result of the automatic alignment}
%\mbox{\scriptsize control where curve 2 was used as the noise source.}
\mbox{\begin{minipage}{0.5\textwidth}
\scriptsize {\bf Figure 3.} Displacement noises. Curve 1 and 2: Combined
displacement of the seismic noises and suspension's resonant motions in Z and X axes respectively.
Curve 3: Relative displacement between CM1 and CM2 in Z axis. Curve 4: Simulation result of the
automatic alignment control where curve 2 was used as the noise source.
\end{minipage}}
\end{wrapfigure}
the incoming laser beam and generate fluctuations of beam intensity and beam spot position.
These fluctuations generate noises and degrade the sensitivity in the ellipticity detection.
Therefore, for long integration time and high sensitivity measurement,
it is necessary to maintain the alignment between cavity mirrors and incoming laser beam.  Many techniques
have been developed by the Gravitational-Wave Community to maintain this alignment automatically.  In our auto-alignment control
scheme, we adopt the Ward technique [12].  Beating the rf phase-modulated TEM$_{00}$ sidebands reflected from the front
cavity mirror and the TEM$_{10}$ of the main band caused by misalignment of cavity mirrors on a split photodiode, we can extract the
misalignment signals between these two beams after rf demodulation.  In figure 1, QPD1 and QPD2 detect this beat signal.
The extracted signals can be linearly combined and fed back to the cavity mirrors via coil-magnet pairs.  A simple
control servo design and closed-loop-performance simulation have been carried out and show that the transversal motion is reduced to below 10$^{-8}$ m/Hz$^{1/2}$ in frequency range between 1-20 Hz (Curve 4 of Figure 3).  In this simulation, the measured seismic noise (Curve 2 of Figure 3) were
used as the noise source.  If necessary, an extra quadrant photodiode will be used to monitor the beam spot position on the analyzer; the position signals
can also be fed back to the peizoelectric driven mirror M3 to stabilize beam spot position further and thus decrease the noise in the detected ellipticity.

\section{Optical Ellipsometer}
Our original optical ellipsometer consists of a Glan-Thomson polarizer and a Glan-Laser analyzer with extinction ratios of the order of 10$^{-7}$, and
a Faraday cell for polarization modulation [8].  We now use two new Glan-Taylor polarizer-analyzer with extinction ratios smaller than 10$^{-8}$
to upgrade our ellipsometer.  We are now performing
a tabletop experiment to test the sensitivity of this optical ellipsometry setup.

\section{Discussion}

In present 2-year goal (Table 1), we hope to increase the sensitivity of ellipsometry by 2 orders of magnitude.  The rotating magnet
is stable up to 10 Hz (cycle per second) with vacuum birefringence modulation up to 20 Hz.  From the measured property of our vibration isolation
system, the noises at these frequencies are 2-order smaller than at 0.05 Hz (previous modulation frequency [8]).  With the implementation of the
automatic alignment control system and the installation of new high-extinction-ratio polarizers, we expect to reach our 2-year goal and to go one step closer
to the detection of vacuum birefringence.

For the next phase after this, with 5-fold improvement on optical sensitivity, 5 m rotation permanent magnet,
and interferometer length extended to 7 m, vacuum birefringence would be in our reach.\\

We thank the National Science Council for supporting this research in part.  We are also grateful to G. Cantatore and C. Rizzo for helpful discussions.

%Figure 1. Schematic setup of the second phase of the Q \& A experiment.

%Figure 2. A picture of the X-pendulum-double-pendulum suspension system.

%Figure 3. Measurement of displacement noises. Curve 1:....................................

\References
\bibitem[1] {Iacopini} Iacopini E and Zavattini E 1979 {\it Phys. Lett.} {\bf 85B} 151
%\bibitem[2] {Ni} Ni W-T \etal 1991 {\it Mod. Phys. Lett.} {\bf A6} 3671
\bibitem[2] {Ni} Ni W-T, Tsubono K, Mio N, Narihara K, Chen S-C, King S-K and Pan S-S 1991 {\it Mod. Phys. Lett.} {\bf A6} 3671
%\bibitem[3] {Cameron} Cameron R \etal 1993 {\it Phys. Rev.} {\bf D47} 3707
\bibitem[3] {Cameron} Cameron R, Cantatore G, Melissinos A C, Ruoso G, Semertzidis Y, Halama H J, Lazarus D M, Prodell A G, Nezrick F, Rizzo C and Zavattini E 1993 {\it Phys. Rev.} {\bf D47} 3707
\bibitem[4] {Pengo} Pengo R \etal 1998
 {\it Frontier Tests of QED and Physics of the Vacuum}  59,
       ed. E Zavattini et al (Sofia: Heron Press); and references therein
\bibitem[5] {Nezrick} Nezrick F 1998
 {\it Frontier Tests of QED and Physics of the Vacuum}  71,
       ed. E Zavattini et al (Sofia: Heron Press); and references therein
\bibitem[6] {Ni} Ni W-T 1998
 {\it Frontier Tests of QED and Physics of the Vacuum}  83,
       ed. E Zavattini et al (Sofia: Heron Press); and references therein

%\bibitem[7] {Zavattini}Zavattini E \etal 2001 {\it Quantum Electrodynamics and
% Physics of the Vacuum} 77, ed. G Cantatore, AIP Conference Proceedings
% Vol.564, American Institute of Physics, New York

\bibitem[7] {Zavattini} Zavattini E, Brandi F, Bregant M, Cantatore S,
Carusotto S, Della Valle F, Di Domenico G, Gastaldi U,
Milotti E, Pengo R, Petrucci G, Polacco E, Ruoso G and Zavattini G 2001 {\it Quantum Electrodynamics and
 Physics of the Vacuum} 77, ed. G Cantatore, AIP Conference Proceedings
 Vol.564, American Institute of Physics, New York

\bibitem[8] {Wu} Wu J-S, Ni W-T and Chen S-J 2003 Building a 3.5 m Prototype Interferometer
      for the Q \& A Vacuum Birefringence Experiment and High Precision ellipsometry,
      paper presented at {\it 5th Amaldi Conf. (Tirrenia, July 2003)}
%\bibitem[9] {Askenazy} Askenazy S \etal 2001 {\it Quantum Electrodynamics and Phisics of the Vacuum} 115, ed. G Cantatore (AIP)
\bibitem[9] {Askenazy} Askenazy S, Billette J, Dupre P, Ganau P, Mackowski J, Marquez J, Pinard L, Portugall O, Ricard D, Rikken G L J A, Rizzo C, Trenec G and Vigue J 2001 {\it Quantum Electrodynamics and Phisics of the Vacuum} 115, ed. G Cantatore (AIP)
\bibitem[10] {Mei} Mei H-H, Ni W-T and Chen S-J 2003 A Heterodyne Interferometry Scheme for Detecting
      the Vacuum Birefringence in in Strong Magnetic Field,
      to be presented at {\it 5th Amaldi Conf. (Tirrenia, July 2003)}
\bibitem[11] {xpen} Tatsumi D, Barton M A, Uchiyama T and Kuroda K 1999 \RSI {\bf 70} 1561; and references therein
%\bibitem[12] {Morrison} Morrison E \etal 1994 {\it Appl. Opt.} {\bf 33} 5037, 5041
\bibitem[12] {Morrison} Morrison E, Beers B J, Robertson D I and Ward H 1994 {\it Appl. Opt.} {\bf 33} 5037, 5041
%\bibitem[13] {Grote} Grote H \etal 2002 {\it Class. Quantum Grav.} {\bf 19} 1849

\endrefs
\end{document}